\documentclass[pra,twocolumn,superscriptaddress,showpacs,floatfix, longbibliography]{revtex4-2}
\usepackage{mathrsfs,braket}
\usepackage{amssymb, amsbsy, amsmath, latexsym, dsfont, array, layout,
graphicx,mathrsfs,color,ulem,bm}
\usepackage[colorlinks=true,citecolor=blue,urlcolor=blue]{hyperref}

\begin{document}

\title{Lee-Yang Zeros of a Bosonic system associated with a single trapped ion}
\author{Wenjie Shao}
\affiliation{CAS Key Laboratory of Microscale Magnetic Resonance and School of Physical Sciences, University of Science and Technology of China, Hefei 230026, China}
\affiliation{CAS Center for Excellence in Quantum Information and Quantum Physics, University of Science and Technology of China, Hefei 230026, China}
\author{Yulian Chen}
\author{Ren-bao Liu}
\affiliation{Department of Physics, The Chinese University of Hong Kong, Hong Kong, China}
\affiliation{Centre for Quantum Coherence, The Chinese University of Hong Kong, Hong Kong, China}
\affiliation{The Hong Kong Institute of Quantum Information Science and Technology, The Chinese University of Hong Kong, Hong Kong, China}
\author{Yiheng Lin}
\email{yiheng@ustc.edu.cn}
\affiliation{CAS Key Laboratory of Microscale Magnetic Resonance and School of Physical Sciences, University of Science and Technology of China, Hefei 230026, China}
\affiliation{CAS Center for Excellence in Quantum Information and Quantum Physics, University of Science and Technology of China, Hefei 230026, China}
\affiliation{Hefei National Laboratory, University of Science and Technology of China, Hefei 230088, China}

\begin{abstract}
Zeros of partition functions, in particular Lee-Yang zeros, in a complex plane provide important information for understanding phase transitions. A recent discovery on the equivalence between the coherence of a central quantum system and the partition function of the environment in the complex plane enabled the experimental study of Lee-Yang zeros, with several pioneering experiments on spin systems. Lee-Yang zeros have not been observed in Bosonic systems. Here we propose an experimental scheme to demonstrate Lee-Yang zeros in Bosonic systems associated with a single trapped ion by introducing strong coupling between the spin and motion degrees of freedom, i.e. beyond the weak coupling Lamb-Dicke regime. Our scheme provides new possibilities for quantum simulation of the thermodynamics of Bosonic systems in the complex plane.
\end{abstract}

\maketitle


\section{\label{introduction}Introduction}

The Lee-Yang zeros (LYZ's) are points on the complex plane where the partition function vanishes\cite{LY}. Considering an Ising model under a complex magnetic field, Lee and Yang showed that the partition function is analytic and all physical properties are determined by its zeros. If the LYZ's approach the real axis, a phase transition is expected with diverging free energy, since the zeros of the partition function are precisely the singularities of the free energy.

The LYZ's has been extended to many interesting spin models\cite{LY_spin_1, LY_spin_2} to study phase transition physics, such as dynamical phase transition\cite{dpt} and critical exponents\cite{cri_exponent}. Despite the intractability for complex physical parameters, Wei and Liu\cite{wei_lee-yang_2012} discovered that the coherence of a probe spin coupled to a many-body system can be mapped directly to the partition function. They proposed a method to observe zeros by measuring the time when such coherence vanishes. Subsequent experiments used liquid-state trimethylphosphite molecules\cite{pengExperimentalObservationLeeYang2015} to first observe the LYZ's. A recent experiment\cite{francisManyBodyThermodynamics2021} measured LYZ's with trapped-ion quantum circuits, providing a scalable method for studying partition functions with near-term devices. Although LYZ's of spin systems are widely studied, they remain elusive in Boson systems\cite{LY_boson_1}. Gnatenko et al\cite{LY_boson_2} demonstrated the relationship between the two-time correlation function and LYZ's. However, measuring such correlation is not straightforward in real experiments.

Phonon modes in trapped-ions are suitable tools to study quantum simulations for Boson systems\cite{caiObservationQuantumPhase2021, meiExperimentalRealizationRabiHubbard2022}. Moreover, recent experiments\cite{out_LD, out_LD2} that harness trapped-ion motion beyond the Lamb-Dicke(LD) regime---where the spatial wavefunction is much smaller than coupling electromagnetic wavelength---provide new possibilities for quantum simulations, in which case the higher-order sidebands cannot be neglected. Moreover, the carrier coupling strength has a non-linear dependence on phonon number $n$, which imprints an $n$-dependent phase factor on spin evolution. As for experimental progress, laser cooling with an LD parameter up to $\eta=1.3$ is demonstrated\cite{out_LD_cooling}, allowing for quantum simulations with deeper coupling between spin and phonon modes.

In this work, we propose an experiment scheme to observe LYZ's of a Boson system using a single trapped ion. We use one phonon mode to construct the Boson system of interest and a pair of Zeeman states as the probe spin. Beyond the LD regime, the coupling strength varies non-linearly with respect to phonon number $n$. We first prepare thermal states with such dependence, and apply a carrier transition outside the LD regime. To generate the target thermal state pertaining to the strong coupling interaction, we extend the averaging method\cite{averaging} to the Boson case and simulate such states with an ensemble of coherent states. This method shows a low infidelity that does not interfere with the pattern of zeros. Finally, we demonstrate that the scheme is robust against experimental noise such as heating and possible spin decoherence.

The work is organized as follows. In Sec.~\ref{section:model} we present our model and implementation of the thermal states. In Sec.~\ref{results}, we describe the spin coherence and partition function measurements with a practical scenario. In Sec.~\ref{imperfections}, we demonstrate the robustness of zeros with known experimental imperfections. The scheme is summarized and discussed in Sec.~\ref{Conclusion}.

\section{\label{section:model}Experimental model and scheme}

Here we consider a Bosonic Hamiltonian $H_S$ in the presence of an external field given by $H_B=hH_I$, where $h=h_R+ih_I$ is complex. The partition function can be written as
\begin{equation}
    Z(\beta, h)=\text{Tr}\exp(-\beta(H_s+h_RH_I)-i\beta h_I H_I)
    \label{Eq:PF}
\end{equation}
where $\beta=1/kT$ is the inverse temperature. The exponential term resembles a thermal state of $H_S+h_R H_I$ under evolution $H_I$ for a duration of $\beta h_I$ when $H_S$ and $H_I$ commute. We could further extract the partition function by coupling the system with a probe spin. We consider a diagonal system Hamiltonian, e.g. $H_S=\omega_m a^{\dagger} a$, where $a(a^{\dagger})$ are annihilation(creation) operators.The commuting criteria suggests a diagonal form $H_I=\sum\xi_n \ket{n}\bra{n}$, which gives the necessary coupling Hamiltonian

\begin{equation}
    H=\Omega S_x\otimes H_I
    \label{Eq:H_couple}
\end{equation}

We use the internal state of a single trapped ion as the probe and a phonon mode with frequency $\omega_m$ as the Bosonic system. To achieve the coupling Hamiltonian, we apply optical transitions on resonance with the spin. The interaction Hamiltonian can be reduced to the following form\cite{bible} with a proper rotating frame,
\begin{equation}
   H^{'}=\Omega S_+ \exp(i[\eta(a\text{e}^{-i\omega_m t}+a^{\dagger}\text{e}^{i\omega_m t})])+h.c.
   \label{Eq:Hint}
\end{equation}
where $\eta=kz_0$ is the Lamb-Dicke parameter, i.e. the ratio between wave function and coupling wavelength. In the sideband-resolved limit where $\omega_m$ is much larger than the Rabi frequency $\Omega$, we can further neglect the oscillating terms and get a Hamiltonian diagonal in the motion subspace,
\begin{equation}
    H^{'}=\Omega S_x\otimes \sum_{n} \frac{\Omega_{n}}{\Omega} |n\rangle\langle n|
    \label{Eq:H_int_reduced}
\end{equation}
where $\Omega_n=\Omega e^{-\eta^2/2}L_n(\eta^2)$ is the Rabi frequency between $|\downarrow,n\rangle \leftrightarrow |\uparrow, n\rangle$ and $L_n$ is the Laguerre polynomial. This yields the coupling interaction~\ref{Eq:H_couple} with $\xi_n=\Omega_n/\Omega$. We initialize our system to the product of a superposition state of $|+\rangle$ and $|-\rangle$, i.e. $|\uparrow\rangle$ and a thermal state,
\begin{equation}
    \rho_0 = |\uparrow\rangle \langle \uparrow|\otimes \frac{\exp(-\beta(H_s+h_RH_I))}{Z_0}
    \label{Eq:rho0}
\end{equation}
where $Z_0=\text{Tr}\exp(-\beta(H_s+h_RH_I))$ is the partition function at zero imaginary field. After the evolution stated above, the normalized partition function $Z/Z_0$ is the off-diagonal element of the probe spin density matrix under $|\pm\rangle$ basis. 

\begin{figure}[ht]
    \hspace*{-0.2cm}
    \includegraphics[width=9cm]{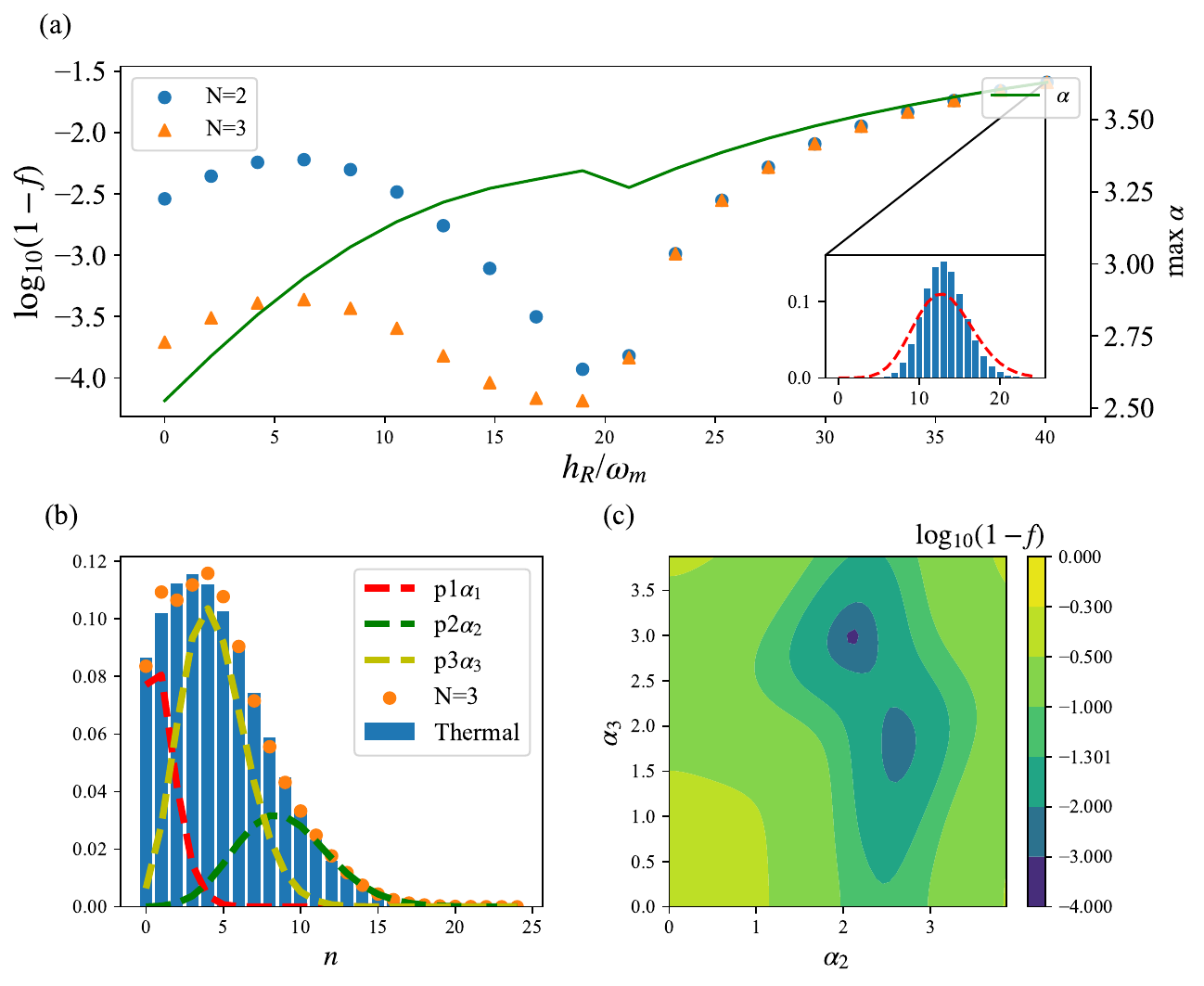}
    \caption{(a)Fidelity of coherent state ensembles under different parameters. The blue circles and orange triangles refer to $N=2$ and 3. The green line shows the maximum displacement $\alpha$ required for preparing an ensemble of $N=3$. In the simulations, we use $\eta=0.47$ and $\beta\omega_m=0.5$. Inset shows an example when the probability distribution(blue bar) becomes sub-Poissonian. Such states cannot be approximated by any coherent state(red dashed line). (b)Thermal states and optimized ensemble at $h_R/\omega_m=7$, which is a typical state exhibiting multiple zeros. (c)Fidelity contour by varying $\alpha_2$ and $\alpha_3$. The fidelity is larger than $99\%$ in the light blue region, which tolerates displacement error by 0.3.}
    \label{fig:fid}
\end{figure}

The probability distribution of our thermal state deviates from any exponential distribution, nor can it be created by any simple operation, such as displacement and squeezing. To generate this density matrix, we prepare a weighted mixture of $N$ coherent states,
\begin{equation}
    \rho_N=\sum_{i=1}^{N}p_i |\alpha_i\rangle \langle \alpha_i|
    \label{Eq:ens}
\end{equation}
where $\alpha_i$ are displacement parameters and their probablities $p_i$ satisfy $\sum p_i=1$. Since the motion subspace is traced out as we measure the spin components, only the diagonal elements of $\rho_N$ will contribute to the spin measurements. Off-diagonal elements of coherent states do not contribute to any observable effect for the probe spin. Therefore we only consider diagonal terms when analyzing the fidelity. Having $N$ fixed, we optimize $p_i$ and $\alpha_i$ to maximize the fidelity $f$. Figure~\ref{fig:fid} shows the infidelity after optimization. By using 3 coherent states the fidelity should reach $99.9\%$, and the maximum displacement in these ensembles is about 3.3, within the range of typical experiments\cite{kienzlerQuantumHarmonicOscillator2015a}. As one increases $h_R/\omega_m$, $H_I$, whose ground-state is a Fock state $\ket{n}$, gradually dominates the thermal distribution. Therefore, our method fails at very large $h_R$ when the target distribution becomes sub-Poissonian. An example of this scenario is shown in the inset of figure~\ref{fig:fid}(a). \\
\indent To examine the robustness of this method against experiment noise, we calculate the fidelity by fixing $\alpha_1$ and vary $\alpha_2$ and $\alpha_3$ for the ensemble shown in figure~\ref{fig:fid}(b). We take a typical parameter $h_R/\omega_m=7$ and $\beta\omega_m=0.5$ that exhibit multiple zeros. As shown by figure~\ref{fig:fid}(c), a fidelity threshold of $99\%$ tolerates fluctuation of $15\%$ on the displacement.

\section{\label{results}Observation of the Partition funtion}
\begin{figure}[h]
    \hspace*{-1cm}
    \includegraphics[width=10cm]{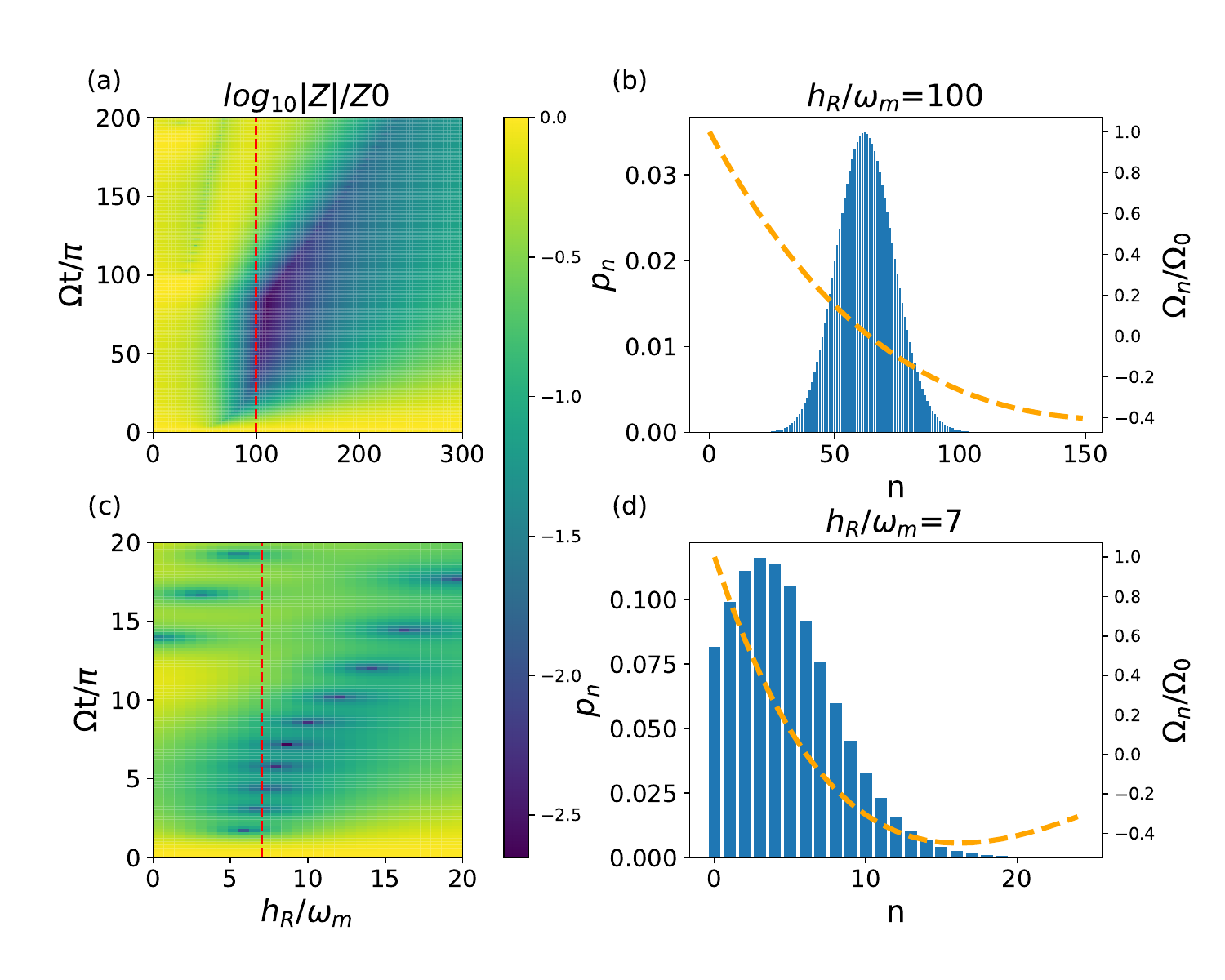}
    \caption{Partition function in the complex plane and corresponding initial distribution in the motion subspace in order to observe LYZ's. The two figures in the left panel show the normalized partition function $Z/Z_0$ with (a)$\eta=0.15$ and (c)$0.47$ respectively. The real axis $h_R/\omega_m$ determines the initial state $\rho_0$ in the motion subspace. Red-dashed lines indicate typical parameters that exhibit LYZ's, of which the probability distributions are shown in (b)(d). Orange dashed lines in (b)(d) are relative Rabi frequencies $\Omega_n/\Omega_0$. The temperature used in these simulations is $\beta\omega_m=0.5$}
    \label{fig:eta}
\end{figure}

Diagonal elements of $H_I$, which depend on $\eta$, will be nearly uniform for $n\leq 20$ if the Lamb-Dicke criterion, i.e. $\eta\ll 1$, is satisfied. This will lead to uniform Rabi oscillation and therefore no spin decoherence. In order to observe Lee-Yang zeros, the initial Gibbs state needs to cover a larger range of Fock states that exhibit variation on Rabi frequencies $\Omega_n$, which is impossible for a precise preparation. Therefore, our experiment scheme favors a larger $\eta$. Figure~\ref{fig:eta} compares Lee-Yang zeros of different $\eta$ at the same temperature $\beta\omega_m=0.5$ and typical initial state probability distributions. When the LD criterion $\eta\ll 1$ is satisfied, one needs to control distributions at $n\simeq 100$ to observe zeros, which is experimentally intractable. Beyond the LD regime, the initial distribution is restricted to $n\leq 20$. This is easier to control for trapped-ion experiments \cite{sensing_oscilator}.

\begin{figure}[h]
    \hspace*{-0cm}
    \includegraphics[width=8.5cm]{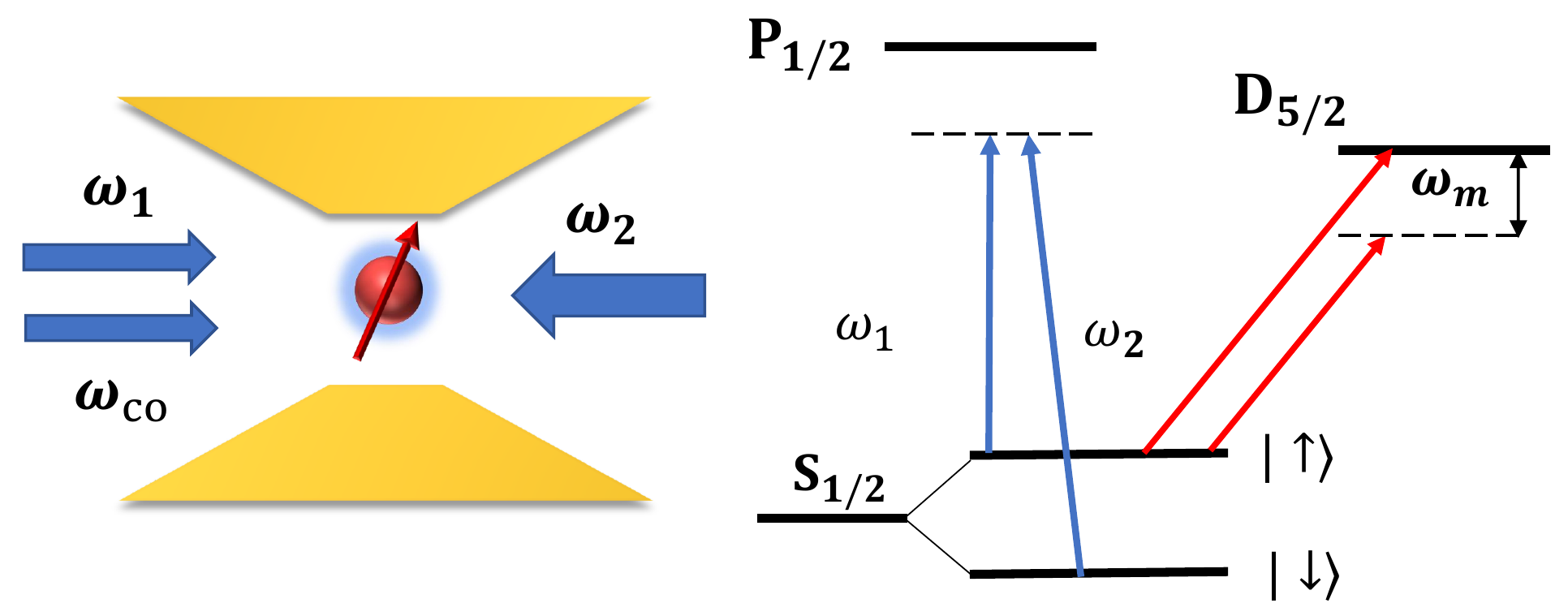}
    \caption{The level scheme of our proposal. The partition function is studied by driving the qubit in the ground-state $^2 S_{1/2}$ manifold. We use a pair of counter-propagating beams along the trap axis at frequency $\omega_1$ and $\omega_2$ to drive Raman transitions. A pair of co-propagating beams $\omega_1$ and $\omega_{cp}$ are used to rotate the spin without Debye-Waller effect. By using a low trap frequency $\omega_m=2\pi\times 600~\text{kHZ}$, we can achieve $\eta=0.47$. The coherent states are prepared with transitions from $^2 S_{1/2}$ to $^{2}D_{5/2}$, as shown by the red arrows.}
    \label{fig:level}
\end{figure}

We consider a practical scenario with a single $^{40}\text{Ca}^+$ ion trapped in the harmonic potential similar to the setup in\cite{poschingerCoherentManipulation402009}. Figure~\ref{fig:level} shows the levels engaged in this scheme. The ground states $|\uparrow\equiv m_s=+1/2\rangle$ and $|\downarrow\equiv m_s=-1/2\rangle$ of $^{2}S_{1/2}$ manifold form the probe spin, and the Boson system we consider is encoded into the phonon mode of the ion. The Zeeman splitting between two states is $2\pi\times20~\text{MHz}$ at a magnetic field of $0.68~\text{mT}$, and the trapping frequency of axial mode is $2\pi\times600~\text{kHz}$. By using a pair of counter-propagating laser beams at $397~\text{nm}$ to drive Raman transitions, we can reach a Lamb-Dicke parameter up to $\eta=0.47$. Similar configurations with large $\eta$ are available with other types of ions such as $^9\text{Be}^{+}$\cite{sensing_oscilator,out_LD_cooling}. A large $\eta$ restricts the motion states of interest to $n\leq20$. We apply laser pulses at $729~\text{nm}$ to couple the $^{2}S_{1/2}$ and $^{2}D_{5/2}$ manifolds. By applying a carrier and a red-sideband(RSB) transition between $|\uparrow\rangle$ and $|^{2}D_{5/2},m_J=1/2\rangle$, the eigenstates are engineered to displaced Fock states\cite{kienzlerQuantumHarmonicOscillator2015a}. In this case, optical pumping from $|^{2}D_{5/2},m_J=1/2\rangle$ to $|\uparrow\rangle$ extracts entropy from the system and eventually cool the ion motion to a displaced ground-state $|\alpha\rangle=D(\alpha)|0\rangle$. By tuning the Rabi frequency ratio between carrier and RSB transition, we can prepare different coherent states. Subsequent summation of the results gives an effective mixed state from different $\alpha$'s 

\begin{figure}[h]
    \hspace*{-0.45cm}
    \includegraphics[width=8.7cm]{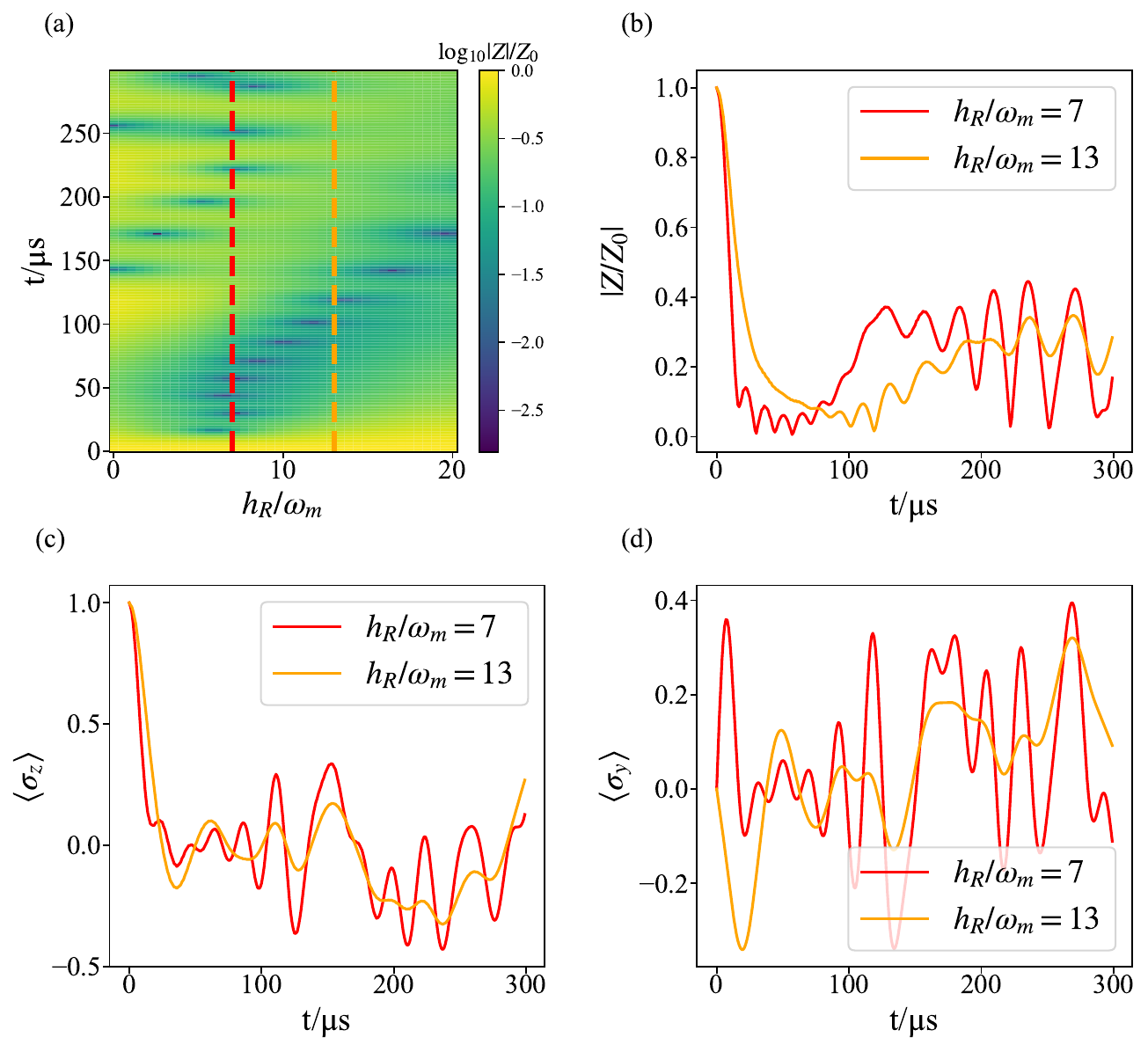}
    \caption{(a)partition function $Z/Z_0$ at $\beta\omega_m=0.5$, Rabi frequency $\Omega=2\pi\times 50~\text{kHz}$. (b)Simulation of spin decoherence on two typical thermal states. The states are labeled as dashed lines in (a). The first state at $h_R/\omega_m=7$ has two groups of zeros, while the second state at $h_R/\omega_m=13$ has one zero in the time of interest.  (c)(d)two spin components, i.e. $\langle\sigma_z\rangle$ and $\langle\sigma_y\rangle$ that we measure in the experiment. A zero partition function requires both components to vanish simultaneously.}
    \label{fig:pf_spin}
\end{figure}

In our scheme, the imaginary axis for LYZ's is mapped to the time of unitary evolution under the interacting Hamiltonian~\ref{Eq:H_int_reduced}. After applying this carrier transition for time $t$, we analyze the spin components perpendicular to $\sigma_x$, i.e. $\sigma_z$ and $\sigma_y$, the latter requires a pair of co-propagating Raman beams to achieve motion-independent rotation $R^{x}_{\pi/2}$
\begin{equation}
    \langle\sigma_z\rangle + i \langle\sigma_y\rangle = \text{Tr}(\rho_N e^{i\Omega H_It})
    \label{Eq:spin}
\end{equation}
We plot in figure~\ref{fig:pf_spin} the partition function and typical spin coherence data that exhibit LYZ's at certain evolution times. By assuming a Rabi frequency of $2\pi\times 50~\text{kHz}$, we found that the spacing between the zeros is above \text{$10~\mu s$}, which can be readily resolved in experiments. 

\begin{figure}[ht]
    \hspace*{-0.1cm}
    \includegraphics[width=9cm]{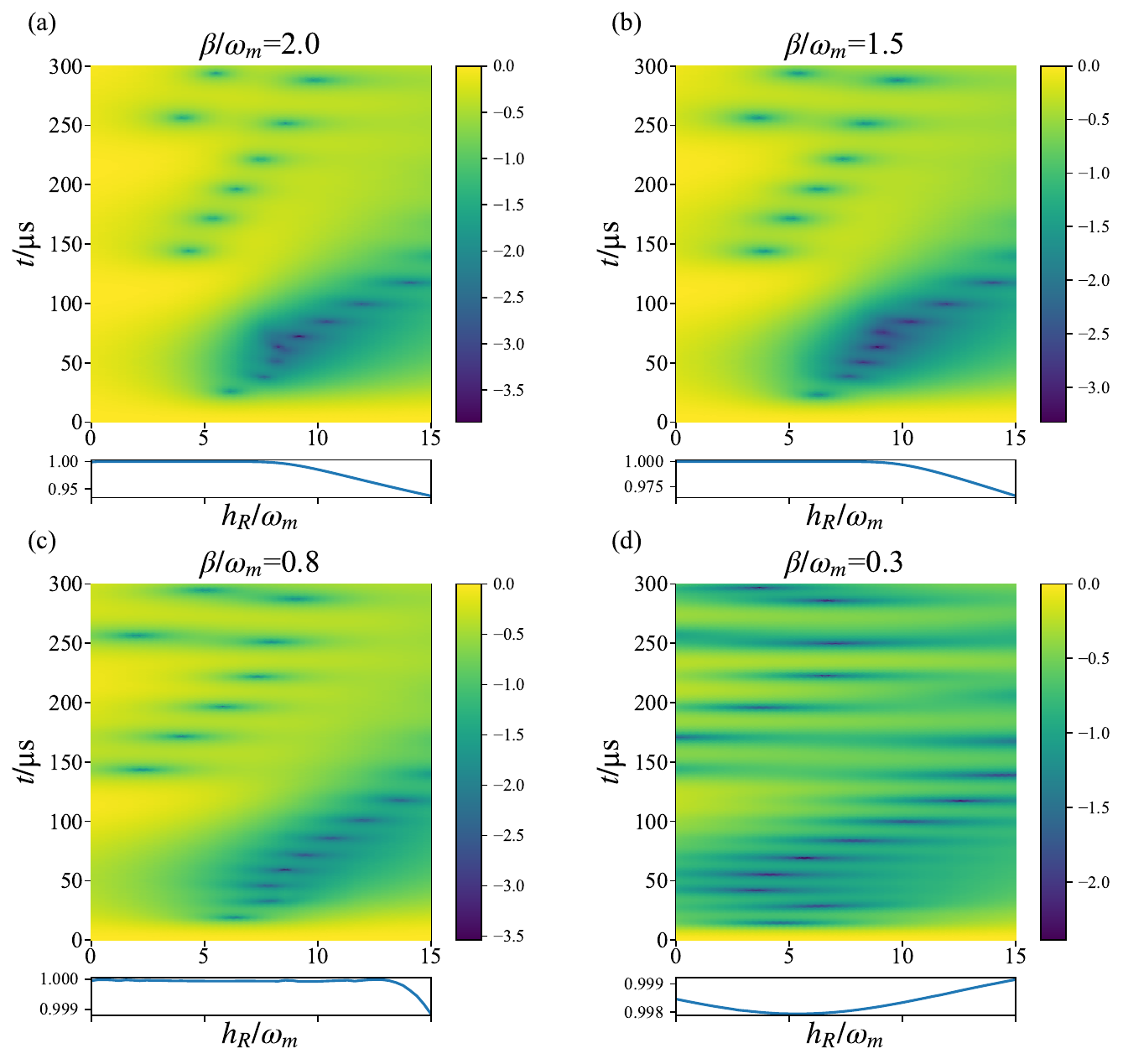}
    \caption{Partition function at different temperatures $\beta\omega_m=$ (a)2.0 (b)1.5 (c)0.8 (d)0.3. Below the contours, we plot the fidelity at each ensemble. At low temperatures(high $\beta$), the target distribution becomes sub-Poissonian and limits the fidelity. At high temperatures (low $\beta$), the low contrast of each zero requires more measurement repetitions.}
    \label{fig:T}
\end{figure}

We also study the partition function at various temperatures to test the generality of our methods. The partition function and fidelity of motional states are plotted in figure~\ref{fig:T}. For lower temperatures(high $\beta$), the sub-Poissonian low entropy thermal states are more difficult to simulate with coherent states, as shown by the decaying fidelity in figure~\ref{fig:T}(a) and (b). For higher temperatures(low $\beta$), the fidelity remains high for the states we consider. However, spin measurement contrasts near zeros are lower as shown in figure~\ref{fig:T}(c). Such a scenario requires more experiment runs for higher resolution.

\section{\label{imperfections}Experimental imperfections}
In this section, we discuss the effect of experimental imperfections on our scheme, and show that our proposal is robust against common imperfections in trapped ion experiments. We first consider the anomalous heating as a noise in the motional subspace. this could be modeled by the Lindblad form master equation, with jump operators $\sqrt{\gamma}a$ and $\sqrt{\gamma}a^{\dagger}$\cite{Gardiner2004QuantumNA},

\begin{equation}
    \frac{\text{d}\rho}{\text{d}t}=-i[H,\rho]+\gamma(a^{\dagger}\rho a-\frac{1}{2}\{aa^{\dagger}, \rho\})+\gamma(a\rho a^{\dagger}-\frac{1}{2}\{a^{\dagger}a, \rho\})
\end{equation}
where $\gamma$ is the heating rate. Since we always start with a product state, the effect of heating is restricted to the motional subspace. In other words, the initial distribution $\rho_N$ we prepare becomes $\rho_N+\Delta \rho$ after heating. We calculate $\Delta\rho$ with the master equation above, and modify the partition function according to equation~\ref{Eq:spin}
\begin{equation}
    \frac{\Delta Z}{Z_0} = \text{Tr}(\Delta\rho e^{i\Omega H_It}) 
    \label{Eq:err_heating}
\end{equation}

We also discuss the influence of spin decoherence. Coherence time of similar systems reach hundreds of microseconds\cite{rusterLonglivedZeemanTrappedion2016a}, about 3 orders of magnitude higer than the time we consider. Besides, noise models that can be described by quantum channels, such as spin dephasing, decay the density matrix exponentially; this should not affect the position of zeros, where the coherence completely vanishes. Therefore, we focus on shot-to-shot frequency fluctuation $V=\frac{1}{2}\Delta\sigma_z$, and calculate the average of experimental runs. A static detuning mixes spin components at $x$ and $z$ direction, and modifies the Rabi frequency by $X_n=\sqrt{(\Delta^2+\Omega_n^2)}$, and the spin measurement outcomes are varied accordingly.
\begin{equation}
    \begin{split}
    \langle\sigma_{z}\rangle^{'} & = \sum_{n}p_{n}(\frac{\Delta^{2}}{X^{2}_{n}} + \frac{\Omega_{n}^{2}}{X_{n}^{2}}\cos{(X_{n} t)}) \\
    \langle\sigma_y\rangle^{'} & = -\sum_{n}p_{n}\frac{\Omega_{n}}{X_{n}}\sin{(X_{n}t)}
    \end{split}
    \label{Eq:spin_detuned}
\end{equation}


By tuning $\eta$, we can maximize the smallest frequency $\Omega_n$ within the range we consider, thus minimizing the term $\Delta^2 / \Omega_n^2$. This yields $\eta=0.455$. 

\begin{figure}[ht]
    \hspace*{-0.25cm}
    \includegraphics[width=9cm]{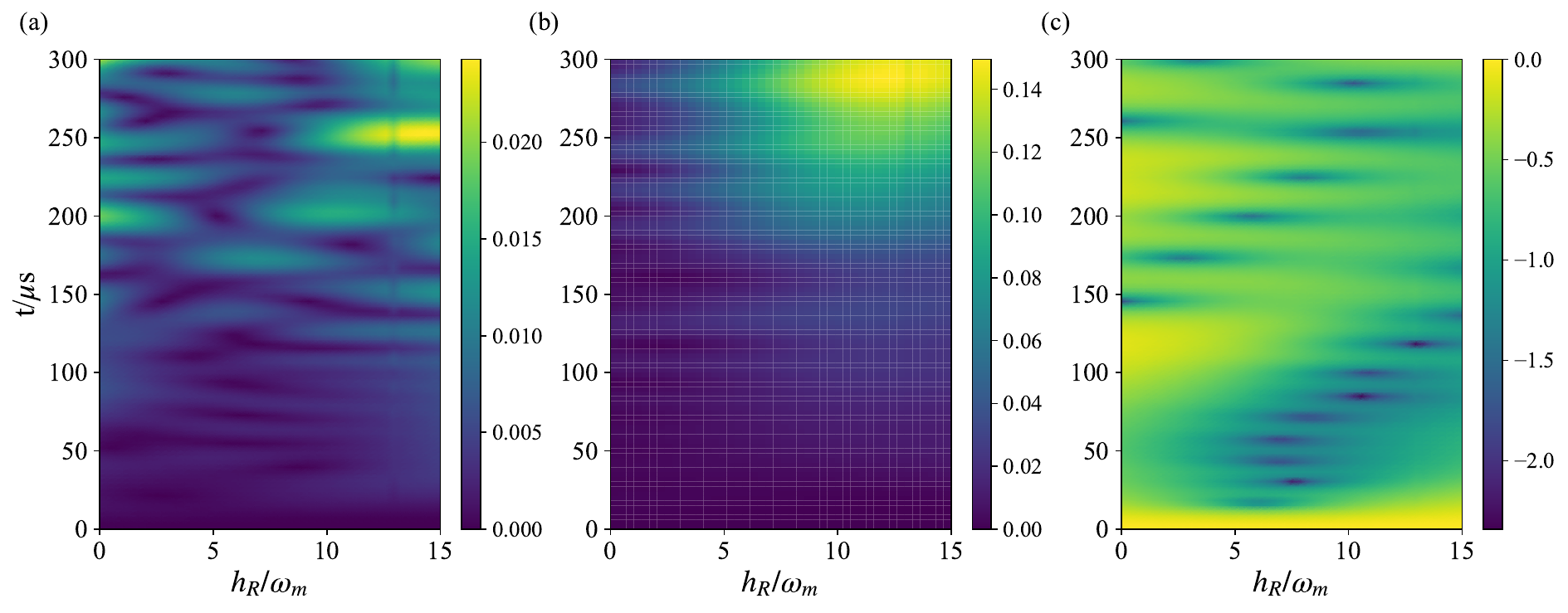}
    \caption{Partition function deviance caused by (a) heating rate of 300 Quanta/s and (b) detuning of 1 kHz, and (c)the partition function with both types of noise. Deviance caused by these two noises does not exceed 0.02 at where the zeros lie. In (c) we use master equations to model the heating rate, and take the average of 100 experiment runs sampling from a Gaussian-distributed shot-to-shot frequency fluctuation to model the spin detuning error. In the simulations we assume $\Omega=2\pi\times 50~\text{kHz}$. }
    \label{fig:noise}
\end{figure}

We calculate the deviance caused by heating and frequency detuning individually according to equation \ref{Eq:err_heating} and \ref{Eq:spin_detuned} in figure~\ref{fig:noise}(a) and (b) by assuming a heating rate of $300~\text{Quanta/s}$ and frequency detuning of $1~\text{kHz}$. Deviations do not exceed $0.02$ at the region of zeros. In figure~\ref{fig:noise}(c) we calculate the partition function with non-zero heating rate and shot-to-shot frequency fluctuations among 100 experiment runs. Compared with figure~\ref{fig:pf_spin}(a), the patterns of zeros are not disturbed. 

\section{\label{Conclusion}Discussion and Outlook}
To summarize, we propose a scheme to simulate LYZ's of a Boson system using a single trapped ion. By applying a carrier interaction beyond the LD regime, we modeled the interaction Hamiltonian that lead to partition function zeros. We showed that the thermal state of a non-linear Hamiltonian can be simulated using an ensemble of coherent states. This method could also be extended to other types of thermal states that cannot be generated with single operations. Furthermore, we proved that our scheme is robust against experimental imperfections. 

For future studies, it is possible to exploit other properties, such as quasi-probability distributions, to explain the pattern of the zeros. Moreover, our method can be scaled up to multiple modes with phase transition, such as the Rabi-Hubbard model.

\begin{acknowledgements}
        \textit{Acknowledgments.---} The USTC team acknowledges support from the National Natural Science Foundation of China (grant number 92165206, 11974330), Innovation Program for Quantum Science and Technology (Grant No. 2021ZD0301603), the USTC start-up funding, and the Fundamental Research Funds for the Central Universities.
\end{acknowledgements}


\begin{thebibliography}{22}%
\makeatletter
\providecommand \@ifxundefined [1]{%
 \@ifx{#1\undefined}
}%
\providecommand \@ifnum [1]{%
 \ifnum #1\expandafter \@firstoftwo
 \else \expandafter \@secondoftwo
 \fi
}%
\providecommand \@ifx [1]{%
 \ifx #1\expandafter \@firstoftwo
 \else \expandafter \@secondoftwo
 \fi
}%
\providecommand \natexlab [1]{#1}%
\providecommand \enquote  [1]{``#1''}%
\providecommand \bibnamefont  [1]{#1}%
\providecommand \bibfnamefont [1]{#1}%
\providecommand \citenamefont [1]{#1}%
\providecommand \href@noop [0]{\@secondoftwo}%
\providecommand \href [0]{\begingroup \@sanitize@url \@href}%
\providecommand \@href[1]{\@@startlink{#1}\@@href}%
\providecommand \@@href[1]{\endgroup#1\@@endlink}%
\providecommand \@sanitize@url [0]{\catcode `\\12\catcode `\$12\catcode
  `\&12\catcode `\#12\catcode `\^12\catcode `\_12\catcode `\%12\relax}%
\providecommand \@@startlink[1]{}%
\providecommand \@@endlink[0]{}%
\providecommand \url  [0]{\begingroup\@sanitize@url \@url }%
\providecommand \@url [1]{\endgroup\@href {#1}{\urlprefix }}%
\providecommand \urlprefix  [0]{URL }%
\providecommand \Eprint [0]{\href }%
\providecommand \doibase [0]{https://doi.org/}%
\providecommand \selectlanguage [0]{\@gobble}%
\providecommand \bibinfo  [0]{\@secondoftwo}%
\providecommand \bibfield  [0]{\@secondoftwo}%
\providecommand \translation [1]{[#1]}%
\providecommand \BibitemOpen [0]{}%
\providecommand \bibitemStop [0]{}%
\providecommand \bibitemNoStop [0]{.\EOS\space}%
\providecommand \EOS [0]{\spacefactor3000\relax}%
\providecommand \BibitemShut  [1]{\csname bibitem#1\endcsname}%
\let\auto@bib@innerbib\@empty
\bibitem [{\citenamefont {Yang}\ and\ \citenamefont {Lee}(1952)}]{LY}%
  \BibitemOpen
  \bibfield  {author} {\bibinfo {author} {\bibfnamefont {C.~N.}\ \bibnamefont
  {Yang}}\ and\ \bibinfo {author} {\bibfnamefont {T.~D.}\ \bibnamefont {Lee}},\
  }\bibfield  {title} {\bibinfo {title} {Statistical theory of equations of
  state and phase transitions. i. theory of condensation},\ }\href
  {https://doi.org/10.1103/PhysRev.87.404} {\bibfield  {journal} {\bibinfo
  {journal} {Phys. Rev.}\ }\textbf {\bibinfo {volume} {87}},\ \bibinfo {pages}
  {404} (\bibinfo {year} {1952})}\BibitemShut {NoStop}%
\bibitem [{\citenamefont {Suzuki}\ and\ \citenamefont
  {Fisher}(2003)}]{LY_spin_1}%
  \BibitemOpen
  \bibfield  {author} {\bibinfo {author} {\bibfnamefont {M.}~\bibnamefont
  {Suzuki}}\ and\ \bibinfo {author} {\bibfnamefont {M.~E.}\ \bibnamefont
  {Fisher}},\ }\bibfield  {title} {\bibinfo {title} {{Zeros of the Partition
  Function for the Heisenberg, Ferroelectric, and General Ising Models}},\
  }\href {https://doi.org/10.1063/1.1665583} {\bibfield  {journal} {\bibinfo
  {journal} {J. Math. Phys.}\ }\textbf {\bibinfo {volume} {12}},\ \bibinfo
  {pages} {235} (\bibinfo {year} {2003})}\BibitemShut {NoStop}%
\bibitem [{\citenamefont {Tong}\ and\ \citenamefont {Liu}(2006)}]{LY_spin_2}%
  \BibitemOpen
  \bibfield  {author} {\bibinfo {author} {\bibfnamefont {P.}~\bibnamefont
  {Tong}}\ and\ \bibinfo {author} {\bibfnamefont {X.}~\bibnamefont {Liu}},\
  }\bibfield  {title} {\bibinfo {title} {Lee-yang zeros of periodic and
  quasiperiodic anisotropic $xy$ chains in a transverse field},\ }\href
  {https://doi.org/10.1103/PhysRevLett.97.017201} {\bibfield  {journal}
  {\bibinfo  {journal} {Phys. Rev. Lett.}\ }\textbf {\bibinfo {volume} {97}},\
  \bibinfo {pages} {017201} (\bibinfo {year} {2006})}\BibitemShut {NoStop}%
\bibitem [{\citenamefont {Heyl}\ \emph {et~al.}(2013)\citenamefont {Heyl},
  \citenamefont {Polkovnikov},\ and\ \citenamefont {Kehrein}}]{dpt}%
  \BibitemOpen
  \bibfield  {author} {\bibinfo {author} {\bibfnamefont {M.}~\bibnamefont
  {Heyl}}, \bibinfo {author} {\bibfnamefont {A.}~\bibnamefont {Polkovnikov}},\
  and\ \bibinfo {author} {\bibfnamefont {S.}~\bibnamefont {Kehrein}},\
  }\bibfield  {title} {\bibinfo {title} {Dynamical quantum phase transitions in
  the transverse-field ising model},\ }\href
  {https://doi.org/10.1103/PhysRevLett.110.135704} {\bibfield  {journal}
  {\bibinfo  {journal} {Phys. Rev. Lett.}\ }\textbf {\bibinfo {volume} {110}},\
  \bibinfo {pages} {135704} (\bibinfo {year} {2013})}\BibitemShut {NoStop}%
\bibitem [{\citenamefont {Deger}\ and\ \citenamefont
  {Flindt}(2019)}]{cri_exponent}%
  \BibitemOpen
  \bibfield  {author} {\bibinfo {author} {\bibfnamefont {A.}~\bibnamefont
  {Deger}}\ and\ \bibinfo {author} {\bibfnamefont {C.}~\bibnamefont {Flindt}},\
  }\bibfield  {title} {\bibinfo {title} {Determination of universal critical
  exponents using lee-yang theory},\ }\href
  {https://doi.org/10.1103/PhysRevResearch.1.023004} {\bibfield  {journal}
  {\bibinfo  {journal} {Phys. Rev. Res.}\ }\textbf {\bibinfo {volume} {1}},\
  \bibinfo {pages} {023004} (\bibinfo {year} {2019})}\BibitemShut {NoStop}%
\bibitem [{\citenamefont {Wei}\ and\ \citenamefont
  {Liu}(2012)}]{wei_lee-yang_2012}%
  \BibitemOpen
  \bibfield  {author} {\bibinfo {author} {\bibfnamefont {B.-B.}\ \bibnamefont
  {Wei}}\ and\ \bibinfo {author} {\bibfnamefont {R.-B.}\ \bibnamefont {Liu}},\
  }\bibfield  {title} {\bibinfo {title} {Lee-{Yang} {Zeros} and {Critical}
  {Times} in {Decoherence} of a {Probe} {Spin} {Coupled} to a {Bath}},\ }\href
  {https://doi.org/10.1103/PhysRevLett.109.185701} {\bibfield  {journal}
  {\bibinfo  {journal} {Phys. Rev. Lett.}\ }\textbf {\bibinfo {volume} {109}},\
  \bibinfo {pages} {185701} (\bibinfo {year} {2012})}\BibitemShut {NoStop}%
\bibitem [{\citenamefont {Peng}\ \emph {et~al.}(2015)\citenamefont {Peng},
  \citenamefont {Zhou}, \citenamefont {Wei}, \citenamefont {Cui}, \citenamefont
  {Du},\ and\ \citenamefont {Liu}}]{pengExperimentalObservationLeeYang2015}%
  \BibitemOpen
  \bibfield  {author} {\bibinfo {author} {\bibfnamefont {X.}~\bibnamefont
  {Peng}}, \bibinfo {author} {\bibfnamefont {H.}~\bibnamefont {Zhou}}, \bibinfo
  {author} {\bibfnamefont {B.-B.}\ \bibnamefont {Wei}}, \bibinfo {author}
  {\bibfnamefont {J.}~\bibnamefont {Cui}}, \bibinfo {author} {\bibfnamefont
  {J.}~\bibnamefont {Du}},\ and\ \bibinfo {author} {\bibfnamefont {R.-B.}\
  \bibnamefont {Liu}},\ }\bibfield  {title} {\bibinfo {title} {Experimental
  observation of lee-yang zeros},\ }\href
  {https://doi.org/10.1103/PhysRevLett.114.010601} {\bibfield  {journal}
  {\bibinfo  {journal} {Phys. Rev. Lett.}\ }\textbf {\bibinfo {volume} {114}},\
  \bibinfo {pages} {010601} (\bibinfo {year} {2015})}\BibitemShut {NoStop}%
\bibitem [{\citenamefont {Francis}\ \emph {et~al.}(2021)\citenamefont
  {Francis}, \citenamefont {Zhu}, \citenamefont {Alderete}, \citenamefont
  {Johri}, \citenamefont {Xiao}, \citenamefont {Freericks}, \citenamefont
  {Monroe}, \citenamefont {Linke},\ and\ \citenamefont
  {Kemper}}]{francisManyBodyThermodynamics2021}%
  \BibitemOpen
  \bibfield  {author} {\bibinfo {author} {\bibfnamefont {A.}~\bibnamefont
  {Francis}}, \bibinfo {author} {\bibfnamefont {D.}~\bibnamefont {Zhu}},
  \bibinfo {author} {\bibfnamefont {C.~H.}\ \bibnamefont {Alderete}}, \bibinfo
  {author} {\bibfnamefont {S.}~\bibnamefont {Johri}}, \bibinfo {author}
  {\bibfnamefont {X.}~\bibnamefont {Xiao}}, \bibinfo {author} {\bibfnamefont
  {J.~K.}\ \bibnamefont {Freericks}}, \bibinfo {author} {\bibfnamefont
  {C.}~\bibnamefont {Monroe}}, \bibinfo {author} {\bibfnamefont {N.~M.}\
  \bibnamefont {Linke}},\ and\ \bibinfo {author} {\bibfnamefont {A.~F.}\
  \bibnamefont {Kemper}},\ }\bibfield  {title} {\bibinfo {title} {Many-body
  thermodynamics on quantum computers via partition function zeros},\ }\href
  {https://doi.org/10.1126/sciadv.abf2447} {\bibfield  {journal} {\bibinfo
  {journal} {Sci. Adv.}\ }\textbf {\bibinfo {volume} {7}},\ \bibinfo {pages}
  {eabf2447} (\bibinfo {year} {2021})}\BibitemShut {NoStop}%
\bibitem [{\citenamefont {M\"ulken}\ \emph {et~al.}(2001)\citenamefont
  {M\"ulken}, \citenamefont {Borrmann}, \citenamefont {Harting},\ and\
  \citenamefont {Stamerjohanns}}]{LY_boson_1}%
  \BibitemOpen
  \bibfield  {author} {\bibinfo {author} {\bibfnamefont {O.}~\bibnamefont
  {M\"ulken}}, \bibinfo {author} {\bibfnamefont {P.}~\bibnamefont {Borrmann}},
  \bibinfo {author} {\bibfnamefont {J.}~\bibnamefont {Harting}},\ and\ \bibinfo
  {author} {\bibfnamefont {H.}~\bibnamefont {Stamerjohanns}},\ }\bibfield
  {title} {\bibinfo {title} {Classification of phase transitions of finite
  bose-einstein condensates in power-law traps by fisher zeros},\ }\href
  {https://doi.org/10.1103/PhysRevA.64.013611} {\bibfield  {journal} {\bibinfo
  {journal} {Phys. Rev. A}\ }\textbf {\bibinfo {volume} {64}},\ \bibinfo
  {pages} {013611} (\bibinfo {year} {2001})}\BibitemShut {NoStop}%
\bibitem [{\citenamefont {Gnatenko}\ \emph {et~al.}(2017)\citenamefont
  {Gnatenko}, \citenamefont {Kargol},\ and\ \citenamefont
  {Tkachuk}}]{LY_boson_2}%
  \BibitemOpen
  \bibfield  {author} {\bibinfo {author} {\bibfnamefont {K.~P.}\ \bibnamefont
  {Gnatenko}}, \bibinfo {author} {\bibfnamefont {A.}~\bibnamefont {Kargol}},\
  and\ \bibinfo {author} {\bibfnamefont {V.~M.}\ \bibnamefont {Tkachuk}},\
  }\bibfield  {title} {\bibinfo {title} {Two-time correlation functions and the
  lee-yang zeros for an interacting bose gas},\ }\href
  {https://doi.org/10.1103/PhysRevE.96.032116} {\bibfield  {journal} {\bibinfo
  {journal} {Phys. Rev. E}\ }\textbf {\bibinfo {volume} {96}},\ \bibinfo
  {pages} {032116} (\bibinfo {year} {2017})}\BibitemShut {NoStop}%
\bibitem [{\citenamefont {Cai}\ \emph {et~al.}(2021)\citenamefont {Cai},
  \citenamefont {Liu}, \citenamefont {Zhao}, \citenamefont {Wu}, \citenamefont
  {Mei}, \citenamefont {Jiang}, \citenamefont {He}, \citenamefont {Zhang},
  \citenamefont {Zhou},\ and\ \citenamefont
  {Duan}}]{caiObservationQuantumPhase2021}%
  \BibitemOpen
  \bibfield  {author} {\bibinfo {author} {\bibfnamefont {M.-L.}\ \bibnamefont
  {Cai}}, \bibinfo {author} {\bibfnamefont {Z.-D.}\ \bibnamefont {Liu}},
  \bibinfo {author} {\bibfnamefont {W.-D.}\ \bibnamefont {Zhao}}, \bibinfo
  {author} {\bibfnamefont {Y.-K.}\ \bibnamefont {Wu}}, \bibinfo {author}
  {\bibfnamefont {Q.-X.}\ \bibnamefont {Mei}}, \bibinfo {author} {\bibfnamefont
  {Y.}~\bibnamefont {Jiang}}, \bibinfo {author} {\bibfnamefont
  {L.}~\bibnamefont {He}}, \bibinfo {author} {\bibfnamefont {X.}~\bibnamefont
  {Zhang}}, \bibinfo {author} {\bibfnamefont {Z.-C.}\ \bibnamefont {Zhou}},\
  and\ \bibinfo {author} {\bibfnamefont {L.-M.}\ \bibnamefont {Duan}},\
  }\bibfield  {title} {\bibinfo {title} {Observation of a quantum phase
  transition in the quantum rabi model with a single trapped ion},\ }\href
  {https://doi.org/10.1038/s41467-021-21425-8} {\bibfield  {journal} {\bibinfo
  {journal} {Nat. Comm.}\ }\textbf {\bibinfo {volume} {12}},\ \bibinfo {pages}
  {1126} (\bibinfo {year} {2021})}\BibitemShut {NoStop}%
\bibitem [{\citenamefont {Mei}\ \emph {et~al.}(2022)\citenamefont {Mei},
  \citenamefont {Li}, \citenamefont {Wu}, \citenamefont {Cai}, \citenamefont
  {Wang}, \citenamefont {Yao}, \citenamefont {Zhou},\ and\ \citenamefont
  {Duan}}]{meiExperimentalRealizationRabiHubbard2022}%
  \BibitemOpen
  \bibfield  {author} {\bibinfo {author} {\bibfnamefont {Q.-X.}\ \bibnamefont
  {Mei}}, \bibinfo {author} {\bibfnamefont {B.-W.}\ \bibnamefont {Li}},
  \bibinfo {author} {\bibfnamefont {Y.-K.}\ \bibnamefont {Wu}}, \bibinfo
  {author} {\bibfnamefont {M.-L.}\ \bibnamefont {Cai}}, \bibinfo {author}
  {\bibfnamefont {Y.}~\bibnamefont {Wang}}, \bibinfo {author} {\bibfnamefont
  {L.}~\bibnamefont {Yao}}, \bibinfo {author} {\bibfnamefont {Z.-C.}\
  \bibnamefont {Zhou}},\ and\ \bibinfo {author} {\bibfnamefont {L.-M.}\
  \bibnamefont {Duan}},\ }\bibfield  {title} {\bibinfo {title} {Experimental
  realization of the rabi-hubbard model with trapped ions},\ }\href
  {https://doi.org/10.1103/PhysRevLett.128.160504} {\bibfield  {journal}
  {\bibinfo  {journal} {Phys. Rev. Lett.}\ }\textbf {\bibinfo {volume} {128}},\
  \bibinfo {pages} {160504} (\bibinfo {year} {2022})}\BibitemShut {NoStop}%
\bibitem [{\citenamefont {Wong-Campos}\ \emph
  {et~al.}(2017{\natexlab{a}})\citenamefont {Wong-Campos}, \citenamefont
  {Moses}, \citenamefont {Johnson},\ and\ \citenamefont {Monroe}}]{out_LD}%
  \BibitemOpen
  \bibfield  {author} {\bibinfo {author} {\bibfnamefont {J.~D.}\ \bibnamefont
  {Wong-Campos}}, \bibinfo {author} {\bibfnamefont {S.~A.}\ \bibnamefont
  {Moses}}, \bibinfo {author} {\bibfnamefont {K.~G.}\ \bibnamefont {Johnson}},\
  and\ \bibinfo {author} {\bibfnamefont {C.}~\bibnamefont {Monroe}},\
  }\bibfield  {title} {\bibinfo {title} {Demonstration of two-atom entanglement
  with ultrafast optical pulses},\ }\href
  {https://doi.org/10.1103/PhysRevLett.119.230501} {\bibfield  {journal}
  {\bibinfo  {journal} {Phys. Rev. Lett.}\ }\textbf {\bibinfo {volume} {119}},\
  \bibinfo {pages} {230501} (\bibinfo {year} {2017}{\natexlab{a}})}\BibitemShut
  {NoStop}%
\bibitem [{\citenamefont {Wong-Campos}\ \emph
  {et~al.}(2017{\natexlab{b}})\citenamefont {Wong-Campos}, \citenamefont
  {Moses}, \citenamefont {Johnson},\ and\ \citenamefont {Monroe}}]{out_LD2}%
  \BibitemOpen
  \bibfield  {author} {\bibinfo {author} {\bibfnamefont {J.~D.}\ \bibnamefont
  {Wong-Campos}}, \bibinfo {author} {\bibfnamefont {S.~A.}\ \bibnamefont
  {Moses}}, \bibinfo {author} {\bibfnamefont {K.~G.}\ \bibnamefont {Johnson}},\
  and\ \bibinfo {author} {\bibfnamefont {C.}~\bibnamefont {Monroe}},\
  }\bibfield  {title} {\bibinfo {title} {Demonstration of two-atom entanglement
  with ultrafast optical pulses},\ }\href
  {https://doi.org/10.1103/PhysRevLett.119.230501} {\bibfield  {journal}
  {\bibinfo  {journal} {Phys. Rev. Lett.}\ }\textbf {\bibinfo {volume} {119}},\
  \bibinfo {pages} {230501} (\bibinfo {year} {2017}{\natexlab{b}})}\BibitemShut
  {NoStop}%
\bibitem [{\citenamefont {Wu}\ \emph {et~al.}(2023)\citenamefont {Wu},
  \citenamefont {Shi},\ and\ \citenamefont {Zhang}}]{out_LD_cooling}%
  \BibitemOpen
  \bibfield  {author} {\bibinfo {author} {\bibfnamefont {Q.}~\bibnamefont
  {Wu}}, \bibinfo {author} {\bibfnamefont {Y.}~\bibnamefont {Shi}},\ and\
  \bibinfo {author} {\bibfnamefont {J.}~\bibnamefont {Zhang}},\ }\bibfield
  {title} {\bibinfo {title} {Continuous raman sideband cooling beyond the
  lamb-dicke regime in a trapped ion chain},\ }\href
  {https://doi.org/10.1103/PhysRevResearch.5.023022} {\bibfield  {journal}
  {\bibinfo  {journal} {Phys. Rev. Res.}\ }\textbf {\bibinfo {volume} {5}},\
  \bibinfo {pages} {023022} (\bibinfo {year} {2023})}\BibitemShut {NoStop}%
\bibitem [{\citenamefont {Knill}\ \emph {et~al.}(1998)\citenamefont {Knill},
  \citenamefont {Chuang},\ and\ \citenamefont {Laflamme}}]{averaging}%
  \BibitemOpen
  \bibfield  {author} {\bibinfo {author} {\bibfnamefont {E.}~\bibnamefont
  {Knill}}, \bibinfo {author} {\bibfnamefont {I.}~\bibnamefont {Chuang}},\ and\
  \bibinfo {author} {\bibfnamefont {R.}~\bibnamefont {Laflamme}},\ }\bibfield
  {title} {\bibinfo {title} {Effective pure states for bulk quantum
  computation},\ }\href {https://doi.org/10.1103/PhysRevA.57.3348} {\bibfield
  {journal} {\bibinfo  {journal} {Phys. Rev. A}\ }\textbf {\bibinfo {volume}
  {57}},\ \bibinfo {pages} {3348} (\bibinfo {year} {1998})}\BibitemShut
  {NoStop}%
\bibitem [{\citenamefont {Wineland}\ \emph {et~al.}(1998)\citenamefont
  {Wineland}, \citenamefont {Monroe}, \citenamefont {Itano}, \citenamefont
  {Leibfried}, \citenamefont {King},\ and\ \citenamefont {Meekhof}}]{bible}%
  \BibitemOpen
  \bibfield  {author} {\bibinfo {author} {\bibfnamefont {D.}~\bibnamefont
  {Wineland}}, \bibinfo {author} {\bibfnamefont {C.}~\bibnamefont {Monroe}},
  \bibinfo {author} {\bibfnamefont {W.}~\bibnamefont {Itano}}, \bibinfo
  {author} {\bibfnamefont {D.}~\bibnamefont {Leibfried}}, \bibinfo {author}
  {\bibfnamefont {B.}~\bibnamefont {King}},\ and\ \bibinfo {author}
  {\bibfnamefont {D.}~\bibnamefont {Meekhof}},\ }\bibfield  {title} {\bibinfo
  {title} {Experimental issues in coherent quantum-state manipulation of
  trapped atomic ions},\ }\href {https://doi.org/10.6028/jres.103.019}
  {\bibfield  {journal} {\bibinfo  {journal} {J. Res. Natl. Inst. Stand.
  Technol.}\ }\textbf {\bibinfo {volume} {103}},\ \bibinfo {pages} {259}
  (\bibinfo {year} {1998})}\BibitemShut {NoStop}%
\bibitem [{\citenamefont {Kienzler}\ \emph {et~al.}(2015)\citenamefont
  {Kienzler}, \citenamefont {Lo}, \citenamefont {Keitch}, \citenamefont
  {de~Clercq}, \citenamefont {Leupold}, \citenamefont {Lindenfelser},
  \citenamefont {Marinelli}, \citenamefont {Negnevitsky},\ and\ \citenamefont
  {Home}}]{kienzlerQuantumHarmonicOscillator2015a}%
  \BibitemOpen
  \bibfield  {author} {\bibinfo {author} {\bibfnamefont {D.}~\bibnamefont
  {Kienzler}}, \bibinfo {author} {\bibfnamefont {H.-Y.}\ \bibnamefont {Lo}},
  \bibinfo {author} {\bibfnamefont {B.}~\bibnamefont {Keitch}}, \bibinfo
  {author} {\bibfnamefont {L.}~\bibnamefont {de~Clercq}}, \bibinfo {author}
  {\bibfnamefont {F.}~\bibnamefont {Leupold}}, \bibinfo {author} {\bibfnamefont
  {F.}~\bibnamefont {Lindenfelser}}, \bibinfo {author} {\bibfnamefont
  {M.}~\bibnamefont {Marinelli}}, \bibinfo {author} {\bibfnamefont
  {V.}~\bibnamefont {Negnevitsky}},\ and\ \bibinfo {author} {\bibfnamefont
  {J.~P.}\ \bibnamefont {Home}},\ }\bibfield  {title} {\bibinfo {title}
  {Quantum harmonic oscillator state synthesis by reservoir engineering},\
  }\href {https://doi.org/10.1126/science.1261033} {\bibfield  {journal}
  {\bibinfo  {journal} {Science}\ }\textbf {\bibinfo {volume} {347}},\ \bibinfo
  {pages} {53} (\bibinfo {year} {2015})}\BibitemShut {NoStop}%
\bibitem [{\citenamefont {McCormick}\ \emph {et~al.}(2019)\citenamefont
  {McCormick}, \citenamefont {Keller}, \citenamefont {Burd}, \citenamefont
  {Wineland}, \citenamefont {Wilson},\ and\ \citenamefont
  {Leibfried}}]{sensing_oscilator}%
  \BibitemOpen
  \bibfield  {author} {\bibinfo {author} {\bibfnamefont {K.~C.}\ \bibnamefont
  {McCormick}}, \bibinfo {author} {\bibfnamefont {J.}~\bibnamefont {Keller}},
  \bibinfo {author} {\bibfnamefont {S.~C.}\ \bibnamefont {Burd}}, \bibinfo
  {author} {\bibfnamefont {D.~J.}\ \bibnamefont {Wineland}}, \bibinfo {author}
  {\bibfnamefont {A.~C.}\ \bibnamefont {Wilson}},\ and\ \bibinfo {author}
  {\bibfnamefont {D.}~\bibnamefont {Leibfried}},\ }\bibfield  {title} {\bibinfo
  {title} {Quantum-enhanced sensing of a single-ion mechanical oscillator},\
  }\href {https://doi.org/10.1038/s41586-019-1421-y} {\bibfield  {journal}
  {\bibinfo  {journal} {Nature}\ }\textbf {\bibinfo {volume} {572}},\ \bibinfo
  {pages} {86} (\bibinfo {year} {2019})}\BibitemShut {NoStop}%
\bibitem [{\citenamefont {Poschinger}\ \emph {et~al.}(2009)\citenamefont
  {Poschinger}, \citenamefont {Huber}, \citenamefont {Ziesel}, \citenamefont
  {Dei\ss}, \citenamefont {Hettrich}, \citenamefont {Schulz}, \citenamefont
  {Singer}, \citenamefont {Poulsen}, \citenamefont {Drewsen}, \citenamefont
  {Hendricks},\ and\ \citenamefont
  {Schmidt-Kaler}}]{poschingerCoherentManipulation402009}%
  \BibitemOpen
  \bibfield  {author} {\bibinfo {author} {\bibfnamefont {U.~G.}\ \bibnamefont
  {Poschinger}}, \bibinfo {author} {\bibfnamefont {G.}~\bibnamefont {Huber}},
  \bibinfo {author} {\bibfnamefont {F.}~\bibnamefont {Ziesel}}, \bibinfo
  {author} {\bibfnamefont {M.}~\bibnamefont {Dei\ss}}, \bibinfo {author}
  {\bibfnamefont {M.}~\bibnamefont {Hettrich}}, \bibinfo {author}
  {\bibfnamefont {S.~A.}\ \bibnamefont {Schulz}}, \bibinfo {author}
  {\bibfnamefont {K.}~\bibnamefont {Singer}}, \bibinfo {author} {\bibfnamefont
  {G.}~\bibnamefont {Poulsen}}, \bibinfo {author} {\bibfnamefont
  {M.}~\bibnamefont {Drewsen}}, \bibinfo {author} {\bibfnamefont {R.~J.}\
  \bibnamefont {Hendricks}},\ and\ \bibinfo {author} {\bibfnamefont
  {F.}~\bibnamefont {Schmidt-Kaler}},\ }\bibfield  {title} {\bibinfo {title}
  {Coherent manipulation of a {\textsuperscript{40}} ca {\textsuperscript{+}}
  spin qubit in a micro ion trap},\ }\href
  {https://doi.org/10.1088/0953-4075/42/15/154013} {\bibfield  {journal}
  {\bibinfo  {journal} {J. Phys. B: At. Mol. Opt. Phys.}\ }\textbf {\bibinfo
  {volume} {42}},\ \bibinfo {pages} {154013} (\bibinfo {year}
  {2009})}\BibitemShut {NoStop}%
\bibitem [{\citenamefont {Gardiner}\ and\ \citenamefont
  {Zoller}(2004)}]{Gardiner2004QuantumNA}%
  \BibitemOpen
  \bibfield  {author} {\bibinfo {author} {\bibfnamefont {C.~W.}\ \bibnamefont
  {Gardiner}}\ and\ \bibinfo {author} {\bibfnamefont {P.}~\bibnamefont
  {Zoller}},\ }\bibfield  {title} {\bibinfo {title} {Quantum noise: A handbook
  of markovian and non-markovian quantum stochastic methods with applications
  to quantum optics}\ }(\bibinfo {year} {2004})\BibitemShut {NoStop}%
\bibitem [{\citenamefont {Ruster}\ \emph {et~al.}(2016)\citenamefont {Ruster},
  \citenamefont {Schmiegelow}, \citenamefont {Kaufmann}, \citenamefont
  {Warschburger}, \citenamefont {Schmidt-Kaler},\ and\ \citenamefont
  {Poschinger}}]{rusterLonglivedZeemanTrappedion2016a}%
  \BibitemOpen
  \bibfield  {author} {\bibinfo {author} {\bibfnamefont {T.}~\bibnamefont
  {Ruster}}, \bibinfo {author} {\bibfnamefont {C.~T.}\ \bibnamefont
  {Schmiegelow}}, \bibinfo {author} {\bibfnamefont {H.}~\bibnamefont
  {Kaufmann}}, \bibinfo {author} {\bibfnamefont {C.}~\bibnamefont
  {Warschburger}}, \bibinfo {author} {\bibfnamefont {F.}~\bibnamefont
  {Schmidt-Kaler}},\ and\ \bibinfo {author} {\bibfnamefont {U.~G.}\
  \bibnamefont {Poschinger}},\ }\bibfield  {title} {\bibinfo {title} {A
  long-lived zeeman trapped-ion qubit},\ }\href
  {https://doi.org/10.1007/s00340-016-6527-4} {\bibfield  {journal} {\bibinfo
  {journal} {Appl. Phys. B}\ }\textbf {\bibinfo {volume} {122}},\ \bibinfo
  {pages} {254} (\bibinfo {year} {2016})}\BibitemShut {NoStop}%
\end{thebibliography}
\end{document}